\documentstyle[psfig,epsf]{mn}

\newbox\grsign \setbox\grsign=\hbox{$>$} \newdimen\grdimen
\grdimen=\ht\grsign
\newbox\simlessbox \newbox\simgreatbox
\setbox\simgreatbox=\hbox{\raise.5ex\hbox{$>$}\llap
     {\lower.5ex\hbox{$\sim$}}}\ht1=\grdimen\dp1=0pt
\setbox\simlessbox=\hbox{\raise.5ex\hbox{$<$}\llap
     {\lower.5ex\hbox{$\sim$}}}\ht2=\grdimen\dp2=0pt

\def\simless{\mathrel{\copy\simlessbox}}

\def\Mv{\mbox{$M_{_V}$}}
\newcommand{\bb}{\bibitem[]{bla}}
\newcommand{\ea}{{et al.}}
\newcommand{\delm}{{$\Delta$\it M}}
\newcommand{\mo}{\relax \ifmmode {M_{0}} \else {$M_{0}$}\fi}
\newcommand{\mpar}{\relax \ifmmode {M_{\rm par}} \else {$M_{\rm par}$}\fi}
\newcommand{\spi}{\relax \ifmmode (\sigma_{\pi}/\pi) \else
{($\sigma_{\pi}/\pi$)}\fi}

\title[The Lutz-Kelker bias in trigonometric parallaxes]{
The Lutz-Kelker bias in trigonometric parallaxes\thanks{Based on
data from the ESA Hipparcos astrometry satellite.} }

\author[Ren\'e D. Oudmaijer,  Martin A.T. Groenewegen \& Hans Schrijver]
{
Ren\'e D. Oudmaijer$^1$, Martin A.T. Groenewegen$^2$ \& Hans Schrijver$^3$ \\
$^1$Blackett Laboratory, Imperial College of Science, 
Technology and Medicine, Prince Consort Road, London SW7 2BZ, U.K. \\
$^2$Max-Planck-Institut f\"ur Astrophysik, 
Karl-Schwarzschild-Stra{\ss}e 1, D-85740 Garching, Germany \\
$^3$SRON, Sorbonnelaan 2, NL-3584 CA Utrecht, The Netherlands \\
}

\date{Accepted 1997.
      Received 1997;
      in original form 1997}

\begin{document}

\maketitle

\begin{abstract}

The theoretical prediction that trigonometric parallaxes suffer from a
statistical effect, has become topical again now that the results of the
Hipparcos satellite have become available.  This statistical effect, the
so-called Lutz-Kelker bias, causes measured parallaxes to be too large. 
This has the implication that inferred distances, and hence inferred
luminosities are too small.  Published analytic calculations of the
Lutz-Kelker bias indicate that the inferred luminosity of an object is,
on average, 30\% too small when the error in the parallax is only
17.5\%.  Yet, this bias has never been determined empirically.  In this
paper we investigate whether there is such a bias by comparing the best
Hipparcos parallaxes which ground-based measurements.  We find that
there is indeed a large bias affecting parallaxes, with an average and
scatter comparable to predictions.  We propose a simple method to
correct for the LK bias, and apply it successfully to a sub-sample of
our stars.  We then analyze the sample of 26 `best' Cepheids used by
Feast \& Catchpole (1997) to derive the zero-point of the
Period-Luminosity relation.  The final result is based on the 20
fundamental mode pulsators and leads to a distance modulus to the Large
Magellanic Cloud - based on Cepheid parallaxes- of 18.56 $\pm$ 0.08,
consistent with previous estimates.

\end{abstract}
 

\section{Introduction}

Although discussed as early as 1953 (Trumpler \& Weaver), Lutz \&
Kelker (1973) were the first to quantify the bias in the absolute
magnitude of a star estimated from its observed trigonometric parallax. 
The principle of the Lutz-Kelker bias (hereafter LK bias) is relatively
easy to understand.  A given parallax, $\pi$ with a measurement error
$\sigma_{\pi}$ yields a distance $d$ with an upper and a lower bound. 
Stars at a smaller distance and stars located further away can both scatter
to the observed distance.  Since there are more stars outside
than inside the distance range -simply because of the different sampled
volumes- more stars from outside the distance range will scatter into
the distance range than those inside.  This effect causes a systematic
bias such that measured parallaxes will on average yield too small
distances. 

The magnitude of the bias can be calculated analytically.  Assuming a
uniform distribution of stars, LK found that the mean correction to the
derived absolute magnitude increases with increasing relative error,
reaching --0.43 mag.  for a 17.5\% error in the parallax.  The
correction itself is considerable, but since we deal with a statistical
process describing the numbers of stars scattering inside and outside
the allowed distance range, the bias is represented by a probability
distribution.  This was investigated by Koen (1992), who calculated the
90\% confidence intervals of the correction for the bias.  He found for
the case of a 17.5\% error in the parallax (almost a 6$\sigma$
detection) the same correction as Lutz \& Kelker, but derived that the
90\% confidence interval ranges from +0.33 to --1.44 mag.  This is to be
compared with the observational error of 0.4 mag.  based only on the
propagation of the error in the parallax.  Such a correction is
remarkable indeed, and bears the consequence that parallax measurements
should be corrected for this effect, or stringent selection criteria in
terms of quality of the data should be taken, before the astrophysical
interpretation of the data can be performed.

Indeed, it appears that the only way to take into account the bias
before deriving astrophysical parameters from parallaxes is to use the
tables by Koen (1992) where the correction is given as function of the
relative error in the parallax, \spi .  However, as has been pointed out
by Smith (1987c), the LK correction to an individual parallax
measurement of a star, or to a sample of stars is valid when no {\it a
priori} additional information, like proper motions or knowledge on the
intrinsic absolute magnitude distribution, is available.  In the case of
a Gaussian or top-hat intrinsic magnitude distribution, Smith
(1987a+b+c; see also Turon Lacarrieu \& Cr\'ez\'e 1977) showed that the
correction is a function of the true absolute magnitude, \mo, the
intrinsic spread in \mo, $\sigma_{\mo}$, the observed parallax $\pi$,
and the error on the observed parallax $\sigma_{\pi}$, and not a simple
function of \spi.  
For a magnitude-limited sample (complete
in parallax to a certain limiting magnitude), this correction,
according to Smith (1987c), approaches zero because the combination of the
Malmquist bias and the LK bias lead to a symmetric error in magnitude.
For other samples, the corrections reach asymptotically
the LK values, which  apply when nothing about \mo\ or $\sigma_{\mo}$ is
known.

Considering the implications of the above and the fact that the
presence of the bias has actually never been established empirically,
we devised a simple empirical test using ground-based and Hipparcos
data.

\begin{figure}
\mbox{\epsfxsize=0.47\textwidth\epsfbox[30 170 510 640]
{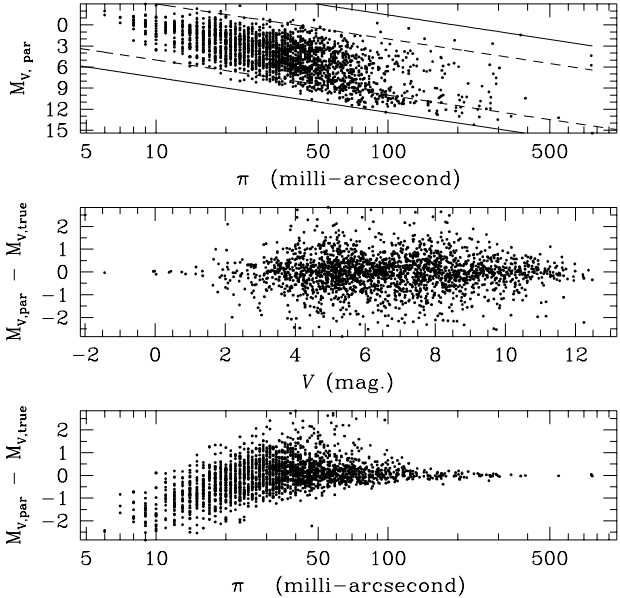}}
\caption{Upper panel: \mpar\ as function of $\pi$.  The ground based
parallaxes are listed with integer values.  The solid lines give an
indication to the zone where we do not expect any data (see text). 
Middle and lower panel: $\Delta M$ as function of {\it V} and  $\pi$.
Although there appears no correlation with {\it V}, there is a strong
correlation with parallax. 
\label{mooi1}
}
\end{figure}

\section{A statistical bias in parallaxes}

We investigate here whether there is any change in the absolute
magnitude determined from the measured parallax $\pi$ of a star as a
function of the error in the parallax \spi. If there would be a trend
towards too faint absolute magnitudes with larger \spi , or even a
large spread, then it is likely that an LK bias is present.

The Hipparcos catalogue is a good starting point to perform such a
test.  It is hard however, to find a sample of stars for which we {\it
know} their absolute magnitudes from first principles.  For example,
selecting stars with the same spectral type will not be sufficient, as
such a sample will have a large spread in intrinsic magnitudes, will
inevitably suffer from misclassifications and have completeness
problems.  On the other hand, when stars with extremely good Hipparcos
parallaxes are considered, one may reasonably assume that their
distances are well determined.  If one then compares such a sample of
stars with their (poorer quality) ground based parallaxes, it is
possible to investigate the LK bias.  We therefore selected all stars
in the Hipparcos Input Catalogue (1993) with trigonometric parallax
measurements 0 $<$ \spi$_{\rm ground-based}$ $<$ 1.  From the
remaining 4007 objects, we selected those stars in the Hipparcos
Catalogue with:

\begin{enumerate}
\item{ 0 $<$  \spi$_{\rm Hip}$ $<$ 0.05, i.e. a 20$\sigma$
detection or better }
\item{ Number of rejected data $<$ 10\% (Field H29)}
\item{ Goodness-of-fit smaller than 3 (Field H30)}
\end{enumerate}

\noindent The first criterion ensures us that we have a sample for
which we may hope to assume that the data do not suffer from
significant bias-problems (if Koen (1992) is correct, 
the mean bias is at most 0.025 mag., with  90\%
confidence limits $\simless$ 0.2 mag.) while the latter
two criteria ensure that the data are not hampered by observational
problems as discussed in the accompanying literature to the Hipparcos
database.  These three extra selection steps left us with a sample of
2187 stars.

\begin{figure}
\mbox{\epsfxsize=0.49\textwidth\epsfbox[30 150 550 500]{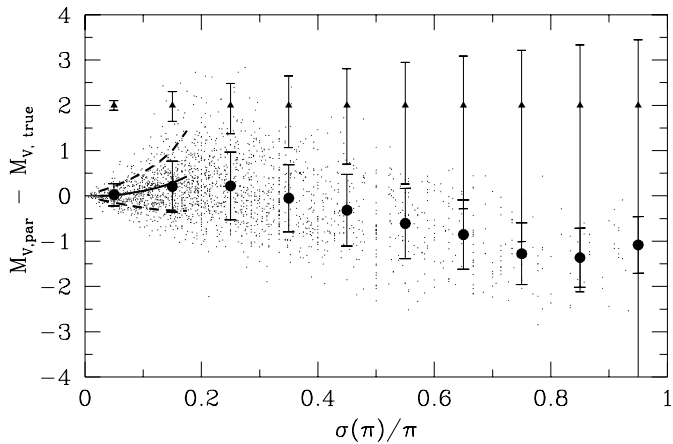}}
\caption{ $\Delta M$ as function of \spi.  The triangles with errorbars
indicate the propagated errors in magnitudes from parallaxes as function
of \spi.  The large solid circles indicate the average $\Delta M$ binned
over intervals of 0.1 in \spi.  
The thick lines indicate the LK bias calculated by
Koen (1992) for a uniform density of stars. The dashed lines indicate the
90\% confidence intervals.
\label{mooi2}
}
\end{figure}

The data are plotted in Figure~\ref{mooi1}.  The upper panel shows the
inferred absolute visual magnitude, derived from the ground-based
parallax, \mpar, plotted against the ground-based parallax.  The stars
follow a well-defined band in the plot.  This is easily understood.
If one takes the analogy for one object with a measured {\it V} band
magnitude, it can only follow a straight line when its derived
intrinsic magnitude is plotted as function of parallax.  Consequently,
all objects should lie between the lines defined by the brightest and
faintest {\it V} magnitude in the sample.  One can also say that the
faintest {\it V} magnitude in the sample defines a minimum possible
value of the parallax for a given value for \mpar.  The difference
between the (measured) absolute magnitude of an object and the
limiting (i.e.  faintest) {\it V} magnitude of a sample implies a
maximum possible distance, and thus minimum parallax.  This can be
written as:

\begin{equation}
\pi > 1000 \times 10^{\left( 0.2\; (M_{\rm parallax} - V_{\rm max} -
5)\right)} \,\,\,\,\,\,\,\,  {\rm mas}
\end{equation}

\noindent
where $V_{\rm max}$ is the faintest magnitude of the sample.  A
similar relation can be written for the brightest (minimum) magnitude
in the sample.  The resulting boundaries are indicated by solid lines
in the upper panel in Fig.~\ref{mooi1}, the dashed lines represent
{\it V} = 2 and 10, encompassing the bulk of the sample.
The middle panel shows the difference (\mpar\ -- \mo), [calculated
from 5\,log($\pi_{\rm Hipparcos} / \pi_{\rm ground-based}$),
hereafter $\Delta M$] as function of {\it V}.  $\Delta M$ shows a
large scatter, especially for faint {\it V}, but the (unweighted) mean
$\Delta M$ = --0.01 $\pm$ 0.8, which would appear rather reassuring.
The propagated errors are not plotted, but these are
indicated in Fig.~\ref{mooi2}.

In the lower panel, $\Delta M$ is plotted against $\pi$.  A strong
correlation is present.  For large $\pi$, $\Delta M$ is close to zero
-indicating good distance determinations-, followed by an increase of
the spread in values until $\Delta M$ decreases towards {\it brighter}
absolute magnitudes.  There are no stars present in the upper left
hand corner of the plot. 
This is not a real effect, but rather a completeness effect in the
data. In reality such objects are  further away than expected,
and too faint to be included in the sample. This zone is
effectively forbidden, as is illustrated by the solid lines in the
upper panel, where the lower left-hand corner is void.

Although completeness effects play an important role,
the resulting distribution of $\Delta M$ is very wide, and illustrates
that absolute magnitude estimates based on trigonometric parallaxes
are subject to large scatter, with a range of --3 to 3 magnitudes.

Figure~\ref{mooi2} shows \delm\ plotted against \spi .  Since the range
in $\sigma_{\pi}$ is much smaller than the range in $\pi$ (the average
$\sigma_{\pi}$ is 7.9 $\pm$ 2.5 mas), the trends are roughly the same as
in the ($\Delta M$ - $\pi$) relation of Figure~1.  The filled circles
indicate the average $\Delta M$, with their standard deviation binnned
over intervals of 0.1 in \spi .  These data are compared with the
results of Koen (1992), in the case of a uniform density of stars and an
infinite number of measurements.  The thick solid line indicates the
mean bias calculated by Koen, and the thick dashed lines represent the
90\% confidence intervals of the bias correction.  For a smaller number
of measurements, the bias and the corrections are larger, while for a
decreasing stellar density (Koen's {\it p} = 2 case), the corrections
themselves are somewhat smaller, but the confidence intervals are
similar.  As most of the stars in our sample are located within 100 pc
(see Fig.~\ref{mooi1}), the assumption of a uniform stellar density is
probably close to the real situation.

For \spi\ $<$ 20 \%, the observations and the predictions by Koen agree
very well, after that, the average $\Delta M$ reaches 0, and goes
towards too bright magnitudes when \spi\ increases further.  It is due
to the objects with large \spi\ (which are affected by the completeness
of the sample) that the overall mean \delm\ is close to zero.  If a
selection on parallax, or \spi , had been applied to these data, the
resulting mean \mpar\ would have led to too faint mean intrinsic
magnitudes.  Unfortunately, our sample renders the construction of a
magnitude limited sample not possible.  It would thus appear that
there is indeed an LK-type bias in parallax data.  In the following, we
will concentrate on the correction for the LK-bias.

\subsection{A correction for the Lutz-Kelker bias}
\label{individual}

One of the assumptions in the analysis by Lutz-Kelker (1973) and Koen
(1992) is that the absolute magnitude and its spread of a sample of
stars are unknown.  Smith (1987b+c) and Turon Lacarrieu \& Cr\'ez\'e
(1977) consider the case of a luminosity function with a Gaussian
spread (for a uniform distribution of stars) and derived a formalism
for the correction of the LK bias.  This correction turns out to be a
function of \mo, $\sigma_{\mo}$, and \spi, and converges to the LK
value for large $\sigma_{\mo}$.  A linear approximation, valid when
$|$\delm $|$ $<$ 2.17, to this rigorous correction was derived by Smith
(1987c):

\begin{equation}
\label{smiteq}
\delta M =  \left[ 1 - \left( \frac{\sigma_{M_0}^{2}}{\sigma_{M_0}^{2}+4.715
\spi^{2}}\right) \right](\mo - \mpar)
\end{equation}

\noindent
For large $\sigma_{M_0}$, the linear approximation becomes zero, 
the exact value of the rigorous correction is close to the
LK value (Smith 1987c).

\begin{figure}
\mbox{\epsfxsize=0.47\textwidth\epsfbox[30 170 510 640]
{
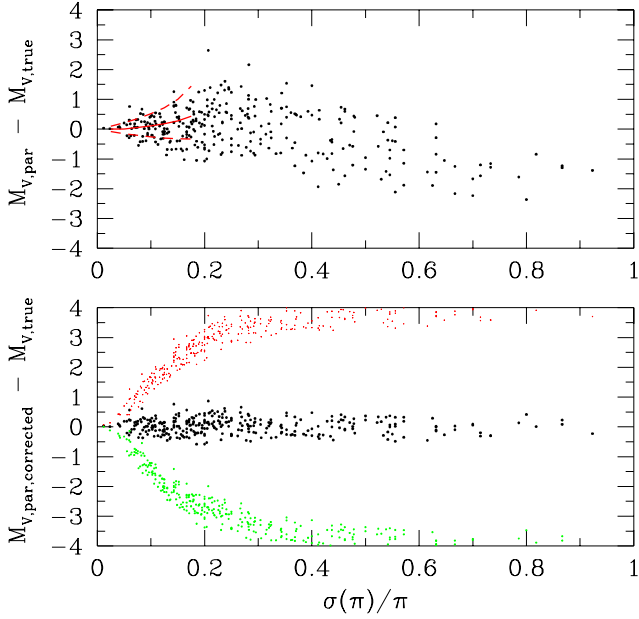}}
\caption{ Upper panel: as the previous figure, 
now for objects with  4 $<$ \mo  $<$ 5. Lower  panel: $\Delta M$ with \mpar\ 
corrected  using Eq.~\ref{smiteq}. The upper and lower clouds of points
respectively show the dependence of the solution when a different \mo\
is used (\mo = 8.5 and 0.5 respectively instead of 4.49). 
\label{smit}
}
\end{figure}

To appreciate the usefulness of this particular result and to see
whether the correction can be used as a means to derive the `true'
intrinsic magnitude of a sample of stars, we select a sub-sample from
our  sample of objects. 
To mimic a sample of stars with approximately the same mean intrinsic
magnitude, we restrict ourselves to a narrow range of true absolute
magnitudes, and apply the correction given in Eq.~\ref{smiteq}.  There
are 313 objects present in the intrinsic magnitude (i.e. derived from
the Hipparcos parallaxes) bin 4 $<$ \mo $<$ 5.
We took the mean \mo, and its standard deviation (4.49 $\pm$ 0.28
mag.) as input values for Eq.~\ref{smiteq}.  The results are plotted
in Fig.~\ref{smit}.  The upper panel shows \delm\ as function of \spi,
which is similar as for the larger sample depicted in
Fig.~\ref{mooi2}.  The corrected values are shown in the lower panel, and
the mean intrinsic magnitude appears to be retrieved.

Hence, it is possible to correct a sample of objects for the LK-bias. 
It would seem that this correction for the statistical biases is not
very useful since one has to know the answer already before applying the
correction.  However, if one studies a sample of objects of which one
may assume that they all have the same \mo, it may be the basis for a
powerful method to derive the absolute magnitude.  To illustrate this,
Fig.~\ref{smit} also shows the resulting \delm\ for other values of \mo\
in Eq.~\ref{smiteq}.  The upper cloud of points was obtained for
inserting \mo\ = 8.5 in the equation, while the lower cloud of points
were calculated using \mo\ = 0.5.  For \spi\ $\simless$ 0.4, a strong
dependence on the input value of \mo\ is present.  It appears then, that
the correct value of \mo\ can be obtained iteratively by varying the
input value of \mo\ to obtain a horizontal line, or a minimum spread
around 0.  Following a procedure outlined later in more detail, we found
the best value for the intrinsic magnitude of the sample to be in the
range 4.48 - 4.59 when small ($<$ 0.1), respectively large ($>$ 0.5),
values of the (assumed to be unknown) spread in intrinsic magnitude are
used.

The above exercise shows that a correction for the LK-bias is not a
simple function of \spi.  In the case of an individual object of which
nothing is known, the LK correction, along with its large confidence
interval is the only remaining option.  However, for a sample of which
all stars are known to have the same intrinsic magnitude, the LK
correction as determined by Smith (1987c) is a potentially powerful tool
to determine \mo.  As shown above, this method returns a surprisingly
good result on our sub-sample of objects.  In the following we will
apply this to a sample of Cepheids. 

\section{The Cepheid distance scale }

Given the presence of a bias in parallax data described above, it is
surprising that several studies based on Hipparcos data, whilst not
taking the LK bias into account, yield results that are relatively close
to previous results.  For example, the new zero-point of the Cepheid
Period-Luminosity (PL) relation that Feast \& Catchpole (1987, hereafter FC)
found, increased the distance modulus to the Large Magellanic Cloud by
(only) 0.2 $\pm$ 0.1 magnitude, and a decrease in the, Cepheid based,
Hubble constant of about 10\%.  Fortunately, we can construct a similar
test for the Cepheids as described above.  Contrary to the sample in the
previous section, where the extremely good Hipparcos parallaxes provided
the true absolute magnitude, we now have to use another indicator.  The
true absolute magnitude of the Cepheids is assumed to follow the PL
relation for Cepheids (cf.  FC):

\begin{equation}
\label{plrel}
<M_{V}> = \delta \ {\rm log}P + \rho
\end{equation}

\noindent
With $\rho$ = --1.43 $\pm$ 0.10 and $\delta$ = --2.81 (as derived and
assumed respectively by FC).  The Hipparcos parallaxes, combined with
the reddening corrected $< ${\it V} $>$ magnitudes then give \mpar.  We
took the data of the 26 `best' Cepheids which contributed the largest
weight (87\% from a sample of 220 objects) to the solutions from FC. 
The stars were corrected for reddening in the same way as in FC. 

\begin{figure}

\centerline{\psfig{figure=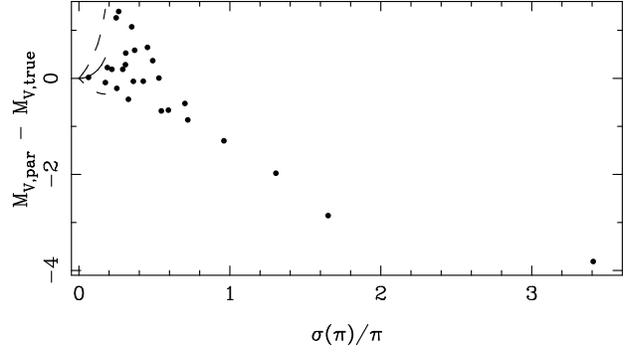,width=8.3cm}}
\caption{As the previous figure, now for Hipparcos measurements of 26
Cepheids.  The true magnitude was calculated using the
Period-Luminosity relation (Eq.~\ref{plrel} with $\rho$ = --1.43).
\label{mooi3}
}
\end{figure}

The relation between \delm\ and \spi\ is plotted in Fig.~\ref{mooi3}. 
The same trend as in Fig.~\ref{mooi2} is present.  For small \spi,
$\Delta M$ tends to indicate fainter intrinsic magnitudes than predicted
from the PL relation, whereas for large \spi, the Cepheids have too
bright magnitudes by as much as 4 mag! The fact that the Cepheids with
large \spi\ are all too bright is due to the completeness effect
discussed earlier.  The  reason that the FC zero-point was close to
previous values is that the LK bias is canceled out by the completeness
effects to a certain, but ill-defined degree.

One can try to correct for the LK-bias using Eq.~\ref{smiteq},
provided we know all relevant parameters.  A difference with the case
illustrated in Fig.~3 is that the parameter we want to estimate is
not $M_{0}$ but $\rho$, the difference between  \Mv\
and $\delta$log{\it P}. A second difference is that  $\sigma_{\rm
M_0}$ (or equivalently  in this case, $\sigma_\rho$) is  unknown.
From the 26 Cepheids we consider the 20 fundamental mode pulsators, as
indicated by FC. For this sample FC found a zero-point $\rho$ = --1.49
$\pm$ 0.13 (corresponding to a distance modulus to the LMC of 18.76).
Three stars have values of $\Delta M$ larger than 2.17, and are hence
not applicable to Eq. 2, and will be discarded in the analysis.
The reason that we chose not to include the six first overtone
pulsators  is discussed below.
We can now  derive an improved value for the zero point
of the Cepheid $PL$ relation applying the LK-correction to the
parallaxes.  

The basic principle behind the following procedure is to iteratively
vary $\rho$ until the variance around \delm\ = 0 is minimal.  We
calculated for a range of values of $\rho$ the `true' \mo\, and
$\rho_{\rm par}$ from $\rho_{\rm par}$ = \mpar\ -- $\delta$ log $P$ for
every star.  The quantity $\Delta \rho$ = ($\rho_{\rm par}$ (corrected)
-- $\rho$) is calculated with $\rho_{\rm par}$ (corrected), the value of
$\rho_{\rm par}$ after applying Eq.  2.  Calculated are the mean and
standard deviation in $\Delta \rho$ and $Q^2$ = $(\sum (\Delta
\rho/s)^2)/(N-1)$.  For $s$ we assumed $\sigma_{\rho}/\sqrt{N}$, with $N
= 17$ the number of stars in the sample.  The best value of $\rho$ is
found where $Q^2$ has a minimum, $\chi^2$.  The 1$\sigma$ uncertainty
around the best value is estimated from those values of $\rho$ for which
$Q^2$ = $\chi^2 + 1$.  
The results are presented
in Table~\ref{hubble2}, where also the standard deviation in the mean of
$\Delta \rho$ is listed. 

A complication is that the spread $\sigma_\rho$ is unknown.  Obviously,
if we use $\sigma_{\rho}$ = 0, all corrections will be
equal to the observed \delm .  All input values of $\rho$ would yield
equally low values of $Q^2$, and hence the error estimate in $\rho$
based on the variation of $\chi^2$ would be very large.  The best value
for $\rho$ is almost equal for every adopted $\sigma_{\rho}$.  We adopt
a final value for $\rho$ = --1.29 $\pm$ 0.02, with the error based on
the scatter in the best fitting values of $\rho$.  This means a decrease
in $\rho$ of 0.20 with respect to FC for {\it exactly} the same sample
and - all things being equal - an LMC distance modulus of 18.56.

\begin{table}
\caption{Zero-point $\rho$ of Cepheids for the 17 best stars
\label{hubble2}    }
\begin{tabular}{llcc}
\hline
\hline
$\sigma_{\rho}$  & $\rho$ & $\sigma$  & $\sigma$ \\
     &  & (from $\chi^2$+1) & (around mean) \\ \hline
0.05     & --1.31  & 2.70  & 0.0017   \\
0.10     & --1.31  & 0.88  & 0.0069 \\
0.175    & --1.30  & 0.55  & 0.020 \\
0.2      & --1.29  & 0.50  & 0.026 \\
0.3      & --1.28  & 0.40  & 0.052 \\ 
0.4      & --1.27  & 0.37  & 0.083 \\
0.6      & --1.25  & 0.35  & 0.143 \\
\hline
\hline
\end{tabular}
\end{table}

\begin{figure}
\centerline{\psfig{figure=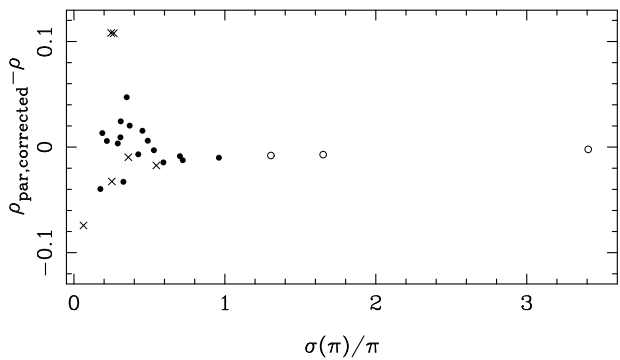,width=8.3cm}}
\caption[]{As previous figure, but now for the fundamental mode
Cepheids, after correction for the LK-bias, for $\rho$ = --1.29. The
open circles represent the three stars not used in the analysis
because they have $\Delta M > 2.17$ (see Fig.~4). The six crosses
represent the first overtone pulsators. Note the change in vertical axis
compared to the previous figure.
}
\end{figure}

The extent to which this procedure has compensated for the bias is
illustrated in Fig.~5, where $\Delta \rho$ is plotted against \spi\
for the case $\rho$ = --1.29, ${\sigma}_{\rho}$ = 0.175 which has a
standard deviation around the mean equal to the adopted uncertainty.
Note that all 20 fundamental pulsators are plotted, although the
fitting was done excluding the three stars with initially the largest
$\Delta M$. The figure also illustrates the reason why we chose not
to include the six first overtone pulsators in the analysis. Three
of them give significantly higher residuals.  This may be taken as
evidence that the procedure by FC to transform first overtone to
fundamental mode pulsation period (Eqs. 8 and 9 in FC) introduces
additional noise.
As FC, we keep the value of $\delta$ fixed. Taking the error into
account will increase the uncertainty in the zero-point. Repeating the
analysis of FC we find for their sample of non-overtone pulsators that
changing $\delta$ by $\pm$0.06 (the uncertainty in the slope derived
by Caldwell \& Laney 1991 and adopted by FC) changes
their zero-point by 0.05. We performed the same test for our
procedure, and find that changing $\delta$ by $\pm$ 0.06 results in a
shift of the best-fitting $\rho$ of $\pm$ 0.05.
Taking into account errors in $\rho$ ($\pm$ 0.02), in $\delta$ ($\pm$
0.05) and in the visual dereddened magnitudes ($\pm$ 0.06) our best
estimate based on a re-analysis of the Cepheid sample of FC is 18.56
$\pm$ 0.08. This value is more in agreement with previous
determinations using Cepheids (see e.g. the listing by Madore \&
Freedman, 1997) and RR Lyrae variables (e.g. Alcock et al. 1997).

\vspace*{-0.5cm}
\section{Concluding remarks}

We have investigated whether there is a Lutz-Kelker type bias present in
parallax data.  For the derivation of the `true' intrinsic magnitudes,
20$\sigma$ parallaxes or better were taken from the Hipparcos data, and
compared with less accurate ground-based parallaxes that were available
for these stars.  For small relative errors in the parallaxes, $\Delta
M$ is distributed a-symmetricly around zero, with a preference for too
faint magnitudes, as is expected from an LK-type  bias.  The spread in values is
consistent with the confidence intervals for the bias that were
calculated by Koen (1992).  For larger values of \spi, the parallaxes
tend to yield too bright magnitudes.  This can be explained by
completeness effects in the data, where the limiting magnitude of the
sample implies a lower limit to the observed parallaxes.  A simple
method to correct for the bias has been presented and tested. 
The Cepheid data of Feast \& Catchpole (1997) were then investigated,
A re-analysis of these data, taking into account any biases, returns a
value of the distance modulus of 18.56 $\pm$ 0.08, which is 0.14
magnitudes smaller than FC found, and in good agreement with previous
determinations.

Finally, we note that unless parallax measurements are extremely
precise, the determination of astrophysical parameters from these data
will be affected by LK-type biases.  For the moment, either using
extremely precise data, or taking into account the LK-bias, with its
large confidence intervals, seems to be the only option for individual
objects, while the simple correction proposed here, can be used to
obtain more reliable estimates of the mean intrinsic magnitude of a
sample of stars, provided one knows beforehand that the objects have the
same luminosity.  The assumption of a uniform distribution of stars used
throughout this paper may  be improved upon by using
Monte-Carlo calculations of the distribution of stars in the line of
sight of the targets.

\paragraph*{Acknowledgments}

We thank Drs. J.E. Drew and L. Lucy for illuminating
discussions. Dr. H. Smith Jr. is thanked for his constructive comments
on an earlier version of the manuscript.  RDO is funded through a PDRA
grant from PPARC.

\end{document}